\documentclass{icrc2009}

\usepackage{graphicx}   
\usepackage[font=footnotesize,caption=false]{subfig} 
\usepackage{fixltx2e}
\usepackage{url}

\newcommand{\shorttitle}[1]%
{\markboth{Proceedings of the 31\MakeLowercase{$^{st}$} ICRC, {\L}\'{o}d\'{z} 2009}{#1} }
\newcommand{\etal}{\MakeLowercase{\textit{et al. }}} 

\begin{document}
\title{Study of electromagnetic backgrounds in the 25-300~MHz frequency band at the South Pole}

\author{\IEEEauthorblockN{Jan Auffenberg\IEEEauthorrefmark{1},
			  									Dave Besson\IEEEauthorrefmark{2},
                          Tom Gaisser\IEEEauthorrefmark{3},
                          Klaus Helbing\IEEEauthorrefmark{1}, 
                          Timo Karg\IEEEauthorrefmark{1},
                          Albrecht Karle\IEEEauthorrefmark{4},\\ and
                          Ilya Kravchenko\IEEEauthorrefmark{5}}
                            \\
\IEEEauthorblockA{\IEEEauthorrefmark{1}Bergische Universit\"{a}t Wuppertal, Fachbereich C- Astroteilchenphysik, 42097 Wuppertal, Germany}
\IEEEauthorblockA{\IEEEauthorrefmark{2}Dept. of Physics \& Astronomy, University of Kansas, Lawrence, KS 66045, USA}
\IEEEauthorblockA{\IEEEauthorrefmark{3}Bartol Research Institute, University of Delaware, Newark, DE 19716, USA}
\IEEEauthorblockA{\IEEEauthorrefmark{4}Dept. of Physics, University of Wisconsin, Madison, WI 53706, USA}
\IEEEauthorblockA{\IEEEauthorrefmark{5}Dept. of Physics \& Astronomy, University of Nebraska, Lincoln, NE 68588, USA}
}

\shorttitle{Jan Auffenberg \etal Electromagnetic background at the South Pole}
\maketitle

\begin{abstract}
Extensive air showers are detectable by radio signals with a radio surface detector. A promising theory of the dominant emission process is the coherent synchrotron radiation emitted by e+ e- shower particles in the Earth's magnetic field (geosynchrotron effect). A radio air~shower detector can extend IceTop, the air shower detector on top of IceCube. This could increase the sensitivity of IceTop to higher shower energies and for inclined showers significantly. Muons from air showers are a major part of the background of the neutrino telescope IceCube. Thus a surface radio air shower detector could act as a veto detector for this muonic background. Initial radio background measurements with a single antenna in 2006 revealed a continuous electromagnetic background promising a low energy threshold of radio air shower detector. However, short pulsed radio interferences can mimic real signals and have to be identified in the frequency range of interest. These properties of the electromagnetic background are being measured at the South Pole during the Antarctic winter 2009 with two different types of surface antennas. In total four antennas are placed at distances ranging up to 400m from each other. They are read out using the RICE DAQ with an amplitude threshold trigger and a minimum bias trigger.
Results of the first three months of measurement are presented.
\end{abstract}

 
\section{Introduction}
\begin{figure}[tbp]
	\includegraphics[angle = -90.0,width=.50\textwidth]{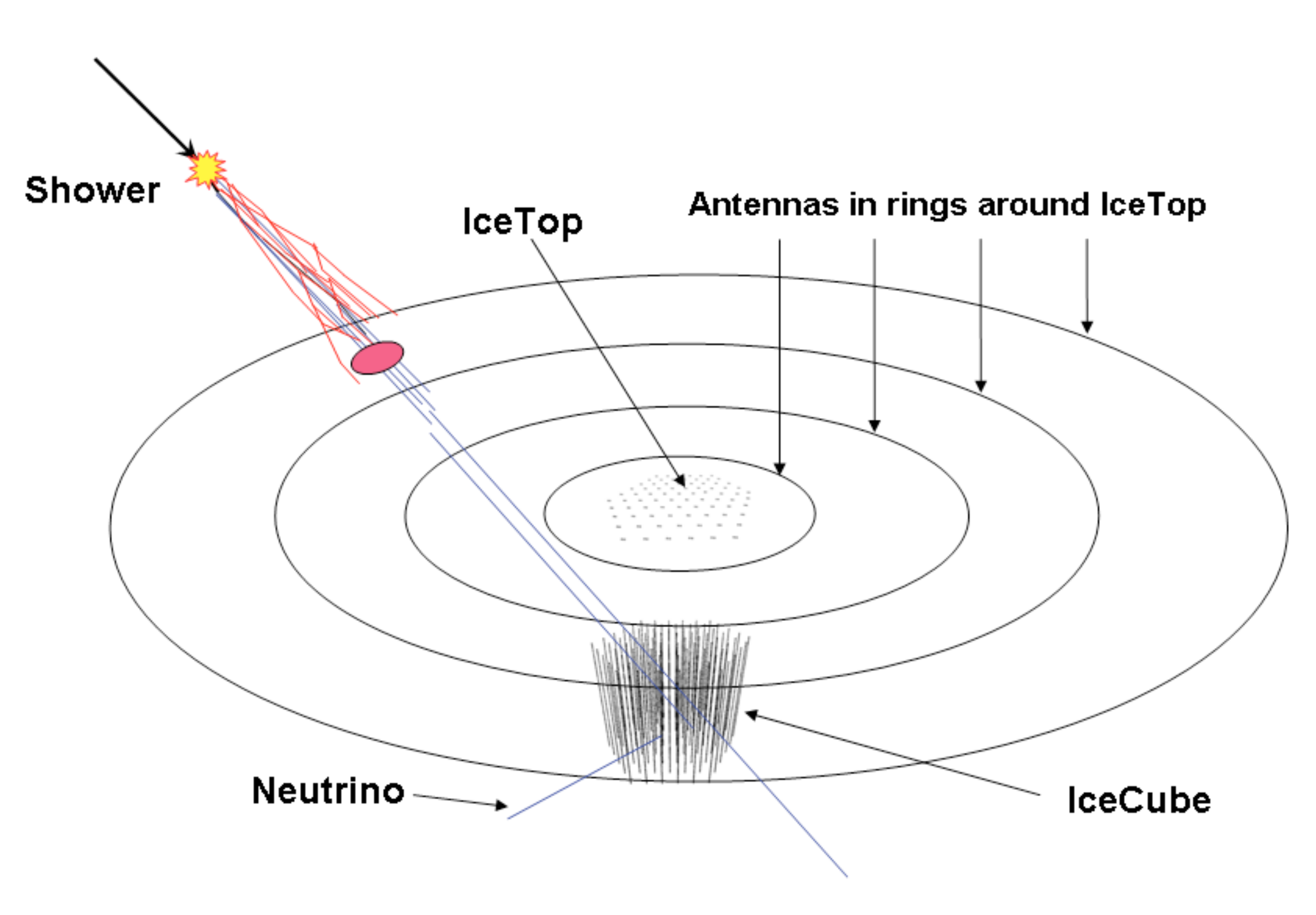}
  \caption{Schematic view of a high energy radio air shower detector
expanding the active area of IceTop. The distance between single
antenna stations can be several hundred meters.}
  \label{fig01}
 \end{figure}

The emission of coherent synchrotron radiation by
$e^+$~$e^-$ shower particles in the Earth magnetic
field provides a measurable broadband signal from 10~MHz-150~MHz on ground \cite{GeoSync}.
The South~Pole site with its dedicated infrastructural environment and a limited number of radio sources is possibly one of the best places in the world for the detection of air~showers by their low frequency radio emission. Another feature of the South Pole site in comparison to other radio quiet regions in the world is the possibility to make studies in coincidence with other astrophysical experiments like the neutrino detector IceCube and the air shower detector IceTop.
IceCube is a neutrino detector \cite{IceCube} embedded in the Antarctic ice. One of the main aims of IceCube is to measure neutrinos from cosmic sources.
The strategy of IceCube is to measure up-going particles from the northern hemisphere. Only neutrinos or other weakly interacting particles are not absorbed by the Earth and are able to interact in the South Pole ice and produce measurable particles like muons. IceTop is built on the surface above IceCube (Fig.~\ref{fig01}). It is designed to detect cosmic air showers
from $10^{15}$~eV up to $10^{18}$~eV.\\
 
This special environment leads to two different options with different focus for a radio air shower detector on top of IceCube and its ambit \cite{ICRC07}.
\begin{itemize}
 \item An \textbf{infill detector} built up of radio antennas on the same footprint as IceTop in similar distances but shifted with respect to the tank array. This provides an additional powerful observation technique in cosmic ray research of air showers at the South~Pole. It would be possible to study air showers by three independent detector systems, IceTop, IceCube and the radio detector.
 \item An \textbf{areal expansion of IceTop} with radio surface antennas is an extension of the air shower detector to higher energy primary particles and to higher inclination angles \cite{Arena08}. The idea is to build an antenna array in rings with increasing radius around the IceTop array. For ultra-high energetic neutrinos, $E>300$~TeV, the neutrino nucleon cross section is large enough for absorption in the Earth to become increasingly important. For cosmogenic neutrinos, produced by the GZK mechanism, for example, most of the signal comes from near the horizon. Thus Muon bundles induced by air showers can be misinterpreted as a neutrino signal in the IceCube detector. The role of an antenna array field expansion of the IceTop detector is to detect these air showers with high inclination angle as a veto for the IceCube detector.
\end{itemize}

\section{Electromagnetic Background Measurements At The South Pole With RICE}

Initial background measurements with a single antenna in 2006 indicated a continuous electromagnetic background promising a low detector threshold \cite{ICRC07}. Together with air shower simulations of the radio emission the background measurements seem to allow the detection of air showers with a threshold lower than 10~PeV in primary energy. The study presented here is aimed at long-term studies of the electromagnetic background for several months to investigate the amount of pulsed radio frequency interference (RFI) and potential long-term variations in the continuous background.
The data acquisition of the RICE experiment, constructed to investigate radio detection methods of high energy neutrinos in ice \cite{RICE}, is suited to be extended by four surface antennas. The RICE DAQ consists of 6 oscilloscopes with 4 channels each. The sampling rate of each channel is 1GHz.
The dynamic range of the scopes is $\pm$2~V with 12~Bit digitization.\\
Three different kinds of trigger are implemented in RICE:

\begin{enumerate}
 \item Unbiased events every 10~min which is a forced read out of all channels.
 \item The RICE simple multiplicity trigger. It is read out if the signal in four or more RICE antennas is above a threshold. The threshold is calculated at the beginning of every run to be 1.5 times above the RMS of unbiased events. These events should have no signal over threshold in the surface antennas.
 \item The RICE surface trigger. This is a RICE simple multiplicity trigger with one or more surface antennas above the threshold as part of the trigger. This includes RICE triggers where only surface antennas have a signal over threshold.
\end{enumerate}
The first and the third kind of trigger are of great interest for the surface radio background studies.
The second trigger strategy is interesting to understand the in-ice RFI not reaching the surface. It is the most interesting event class for the RICE neutrino detection.\\

\begin{figure}[tbp]
	\includegraphics[width=.55\textwidth]{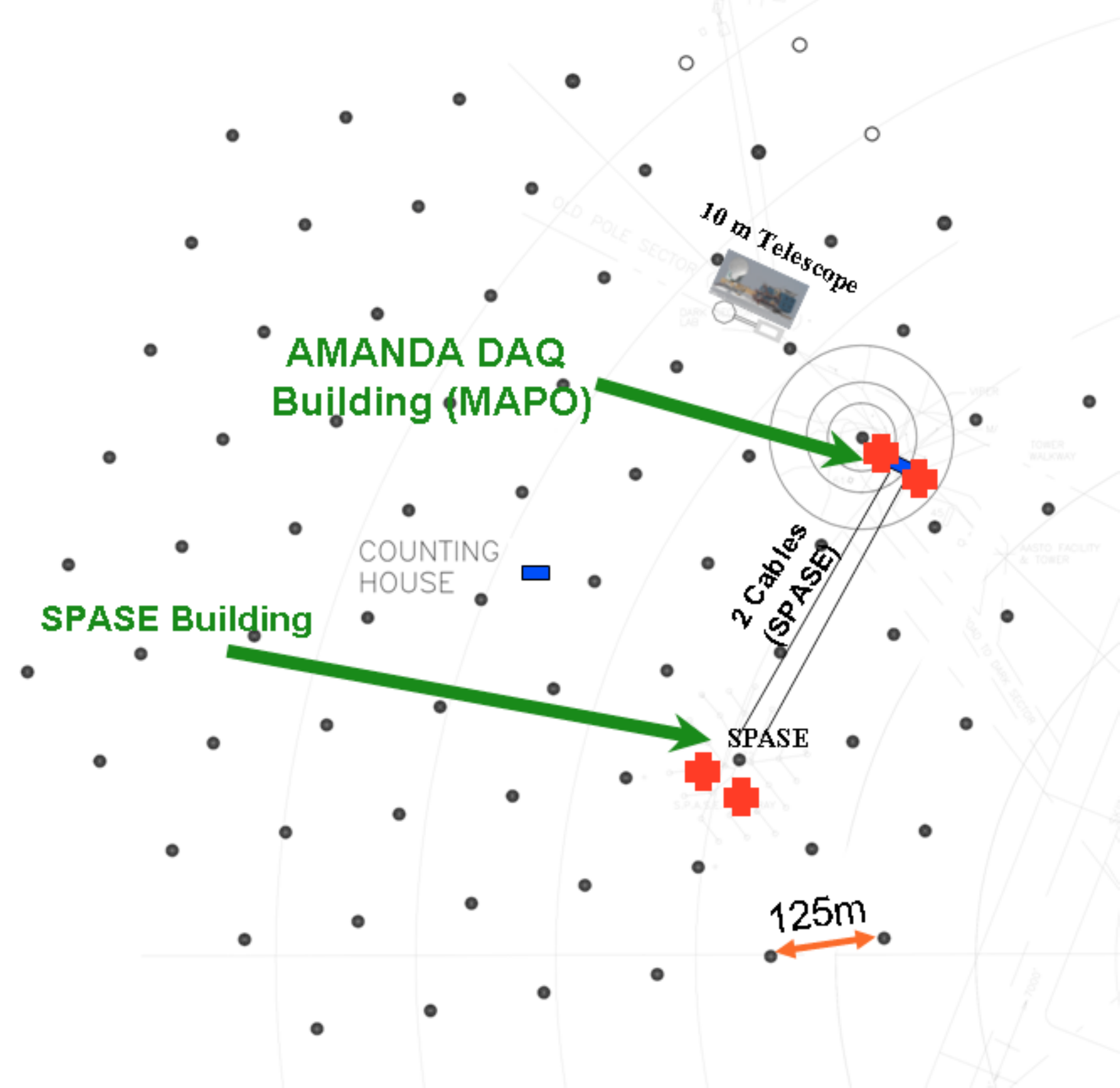}
  \caption{Top view of the IceCube footprint. Two Fat Wire Dipole antennas were deployed in 350~m distance from the MAPO building (crosses near the SPASE building). The signals are amplified with 60dB MITEQ AU-4A-0150 low noise amplifiers and connected with RG59 cables to the MAPO building. Two four arm dipole antennas are located on the roof of the MAPO building and amplified with 39~dB MITEQ AU-1464-400.}
  \label{fig03}
 \end{figure}

In total four surface antennas were deployed in the South Pole season 2008/2009 on the IceCube footprint (Fig.~\ref{fig03}).
\begin{figure}[tbp]
	\includegraphics[width=.50\textwidth]{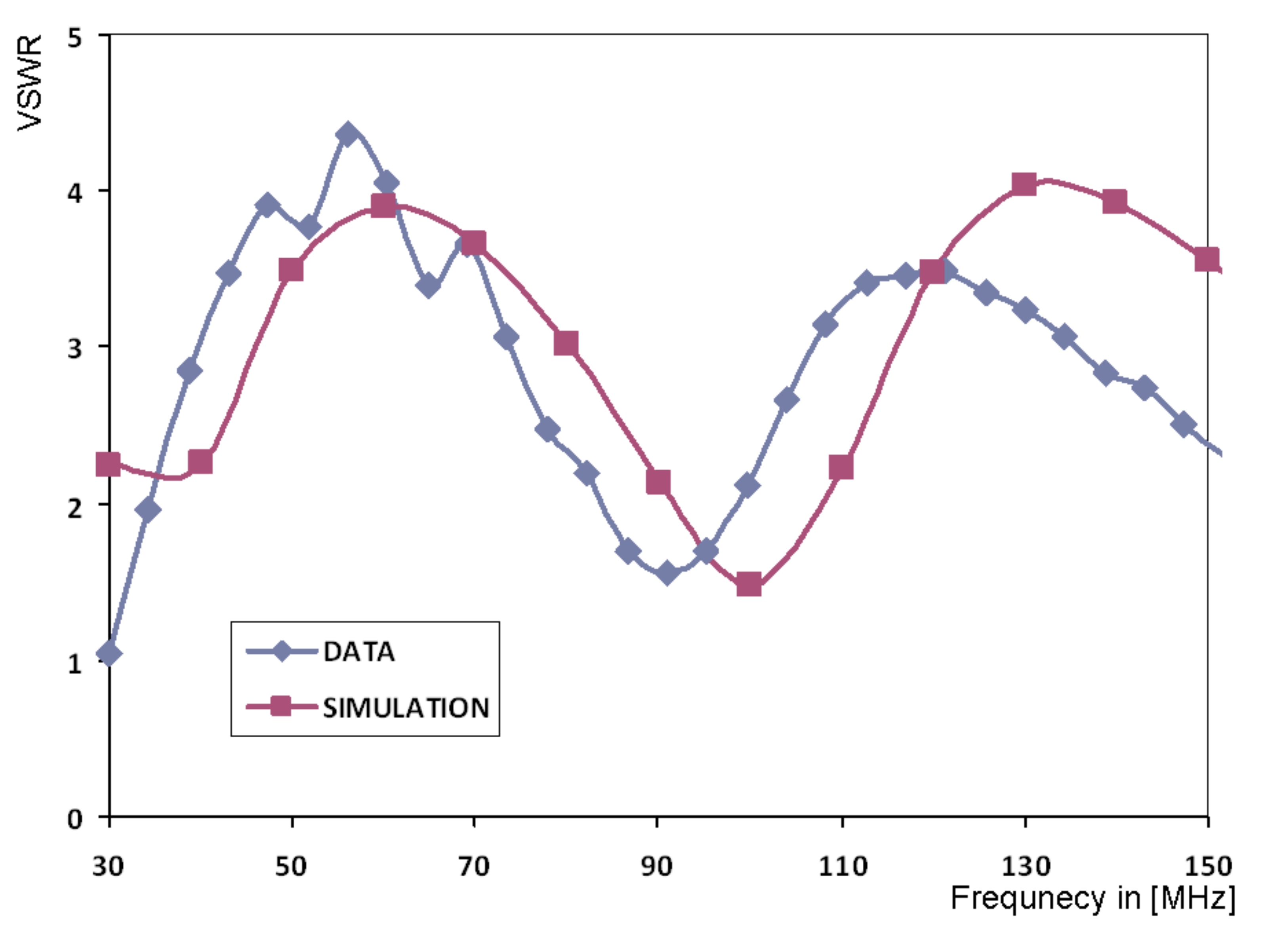}
  \caption{Comparison of results from antenna simulations and measured properties (DATA). The data is a measurement of the voltage standing wave ratio (VSWR) of the Fat Wire Dipole on the South Pole snow at the final position of the antenna. The Simulation is made with EZNEC+ v. 5.0 without ground effects from the snow surface. The frequency response of the antenna is well described by the simulation. Including ground effects should even improve the agreement between simulation and data.}
  \label{fig02}
 \end{figure}
Two Fat Wire Dipole antennas (Fig.~\ref{fig05}) are connected to RICE with RG59 signal cables (1505A Coax) of the decommissioned SPASE experiment (Fig.~\ref{fig03}). These broad band antennas allow for measurements in the frequency range from 25-500~MHz. Figure~\ref{fig02} shows measurements of the voltage standing wave ratio (VSWR) of a Fat Wire Dipole lying on the Antarctic snow in its final position compared to simulations of the antenna using the NEC2 software package \cite{NEC2}. The frequency response is already well described by simulations without snow ground. The connection with the SPASE cables allows measurements over distances of several hundred meters from the MAPO building on the IceCube footprint, housing electronics, which is a potentially large RFI source, and the antennas on its roof. To compensate for the attenuation of the long signal cable (ca. 40~dB at 75~MHz) 60dB low noise preamplifiers (MITEQ AU-4A-0150) are implemented between antenna and signal cable. The power is transmitted through the same cable using bias tees. To avoid saturation of the preamplifiers by the input power, 25~MHz high pass filters and 300~MHz low pass filters were implemented between amplifiers and Fat Wire Dipole antenna. The antennas near the SPASE building are lying on the snow surface orthogonal to each other. Thus they measure orthogonal polarization of the signals.
The other two antennas are deployed on the roof of the MAPO building. These four arm dipole antennas with an amplification of about -2~dB at $>$70~MHz are difficult to calibrate in the surrounding of the MAPO building and thus will only be used for event reconstruction (Fig.~\ref{fig04}).
The signal of the antennas on the roof of the MAPO building are amplified with 39~dB preamplifiers (AU-1464-400). 300~MHz low pass Filters are used for these roof antennas, too. Table \ref{Table1} shows the position of the surface antennas in AMANDA coordinates and their cable delays.

\begin{table}
	\centering
	\caption{Antenna position and cable delays to the RICE DAQ.}
	\begin{tabular}{|c|c|c|c|c|}
  \hline
   Antenna & x [m]  &  y [m] & z [m] & cable delay [ns]\\
   \hline 
   MAPO1 & 47 & -28.0 & 18 & 144\\
   MAPO2 & 25 & -20 & 18 & 129\\
   SPASE1 & -135 & -366 & 1 & 2126\\
   SPASE2 & -148 & -348 & 1 & 2729\\
  \hline
  \end{tabular}
  \label{Table1}
\end{table}

\begin{figure}[tbp]
	\includegraphics[angle = -90.0,width=.51\textwidth]{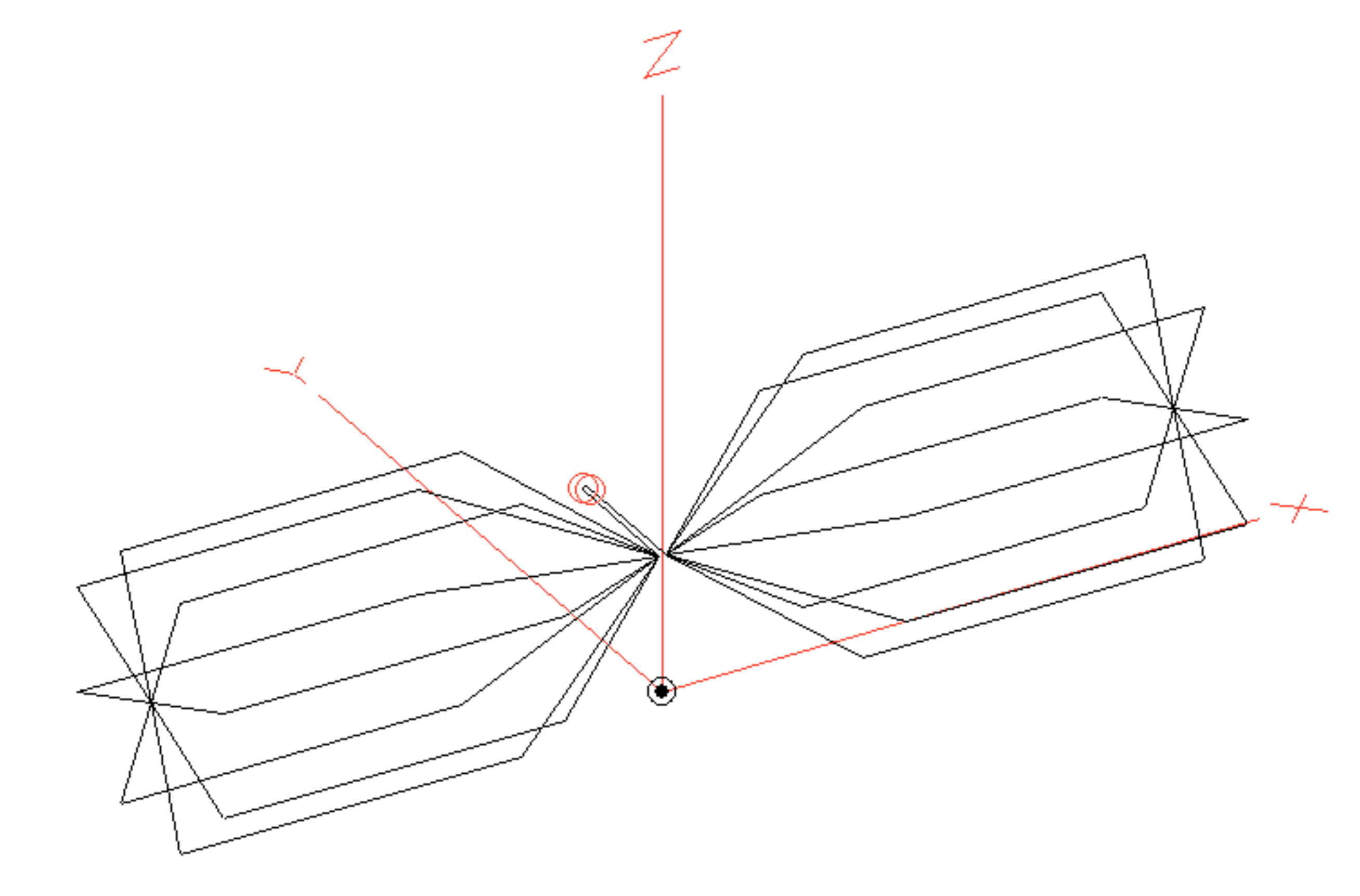}
  \caption{Conducting elements of the Fat Wire Dipole Antenna deployed near the SPASE building. It is 3~m long and 0.8~m in diameter. The wires are mounted to a wooden carcass.}
  \label{fig05}
 \end{figure}

\begin{figure}[tbp]
	\includegraphics[angle = -90.0,width=.49\textwidth]{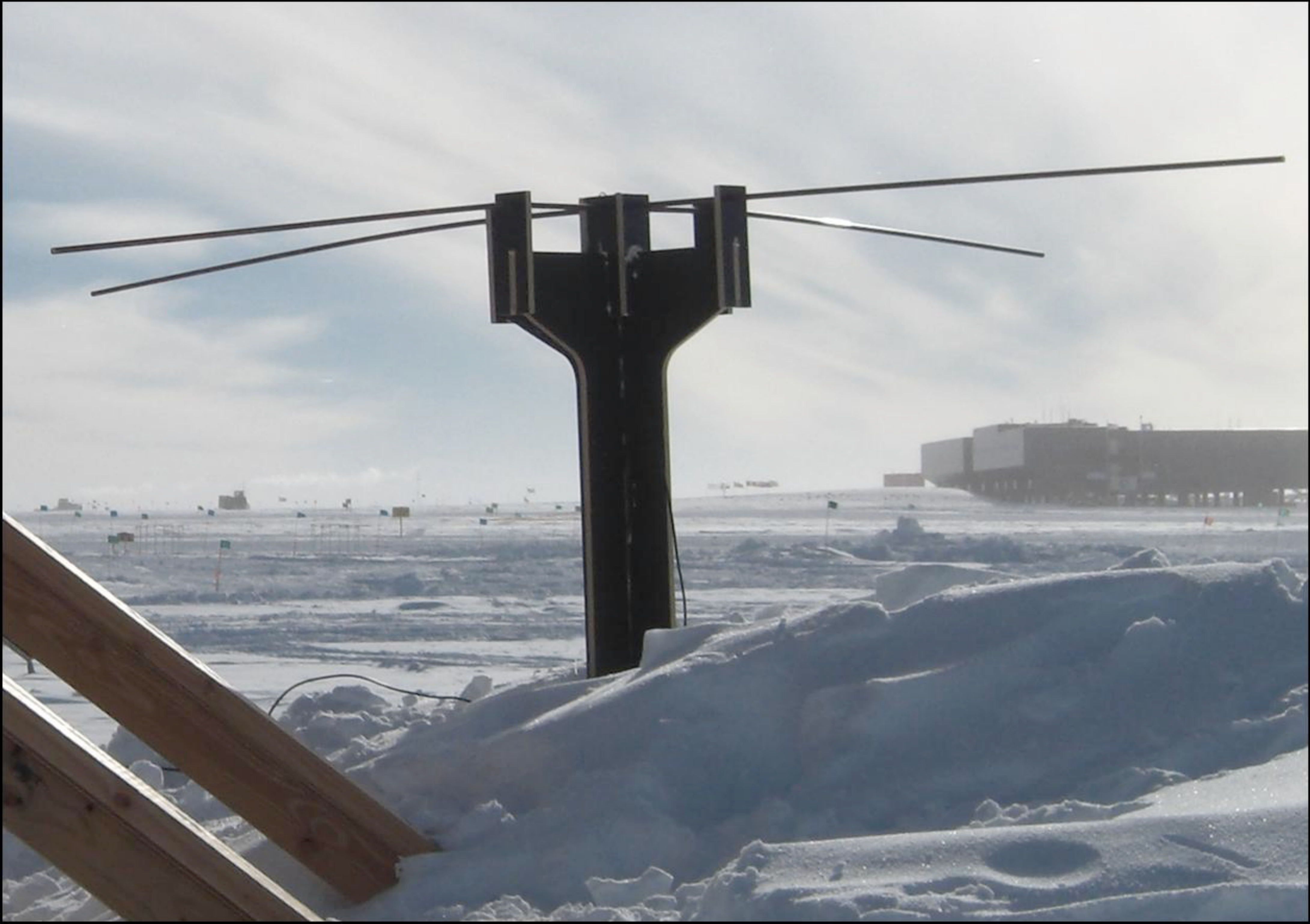}
  \caption{Picture of the four arm dipole antennas MAPO1 and MAPO2. Every arm has a length of 0.7~m. The wooden stand is 1.2~m high. The final position of the antennas is the roof of the MAPO building.}
  \label{fig07}
 \end{figure}

\section{Analysis Strategies}
From the technical point of view one can  divide the analysis of the data into two parts. 
  
\subsection{Event Reconstruction And Mapping}
\begin{figure}[tbp]
	\includegraphics[width=.50\textwidth]{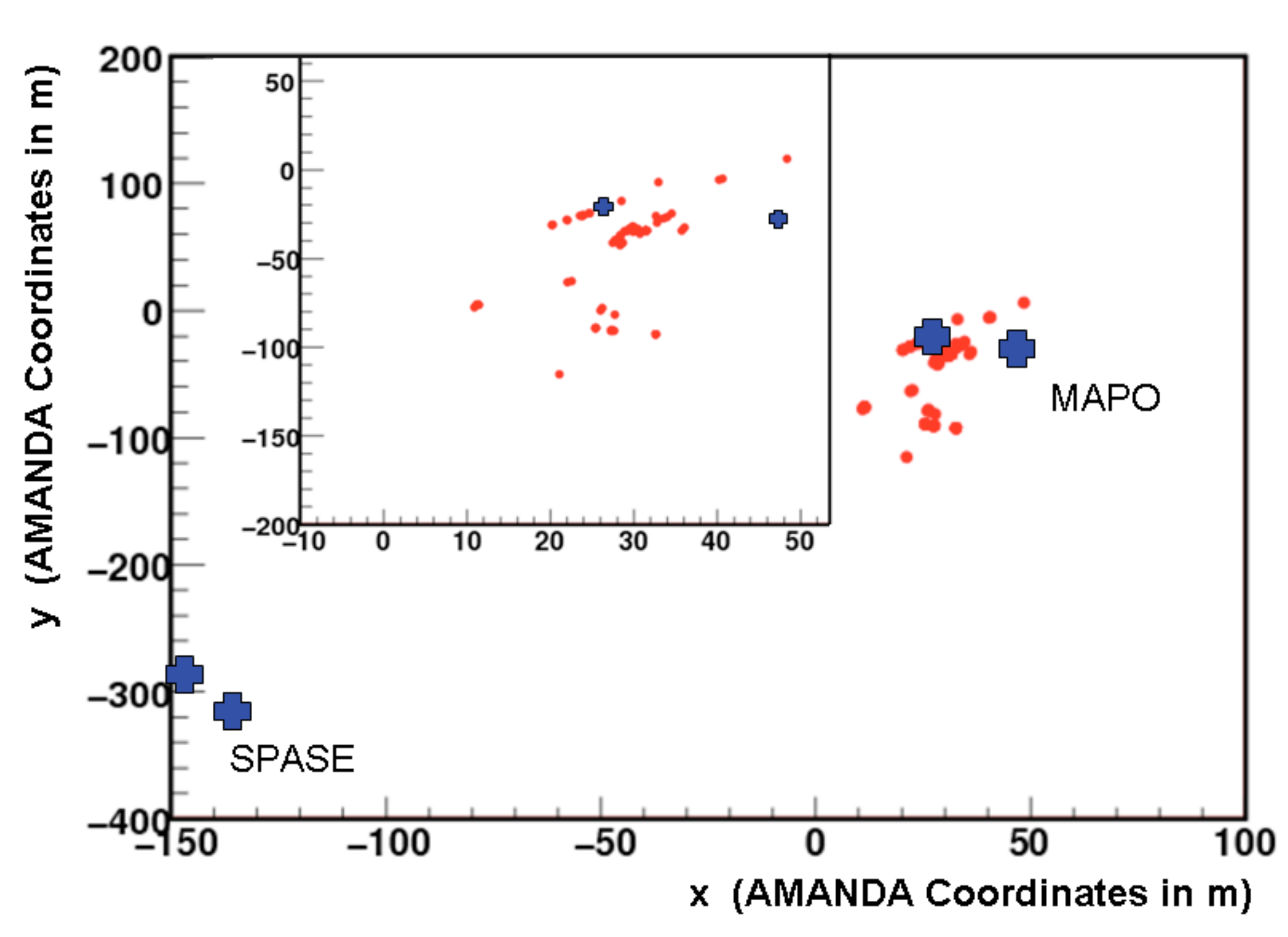}
  \caption{Distribution of the reconstructed transmitter events in the $xy$-plane. The four crosses indicate the position of the antennas on the roof of the MAPO building (18m above the snow) and on the snow surface near the SPASE building. The dots are reconstructed positions of triggered signals from a GHz horn antenna, measured with all four antennas. The reconstructed events are in very good agreement to the transmitter position in front of the MAPO1 antenna and demonstrate the potential of the instrument.}
  \label{fig04}
 \end{figure}

\begin{figure}[tbp]
 \includegraphics[width=.50\textwidth]{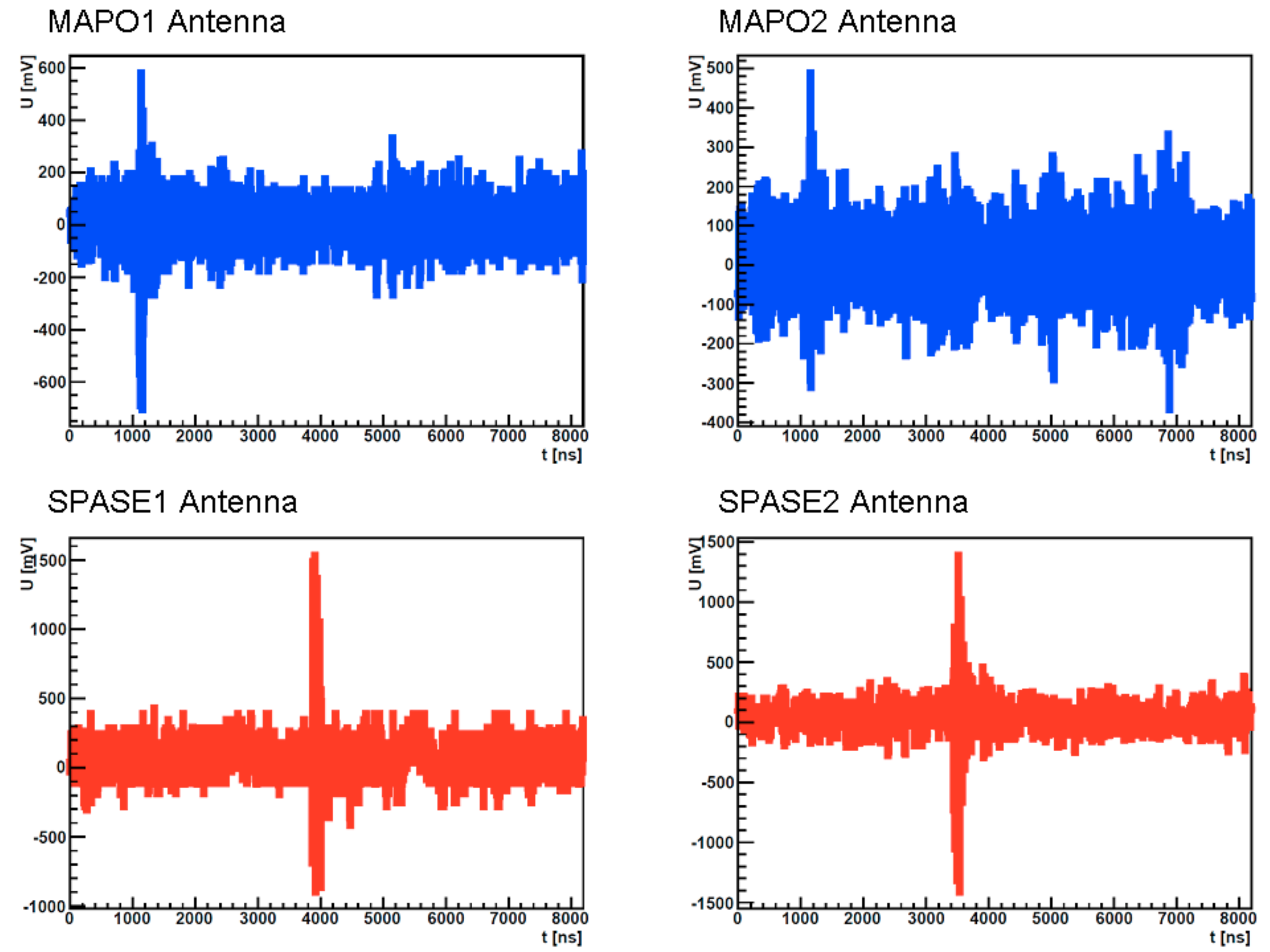}
 \caption{Example of a triggerd event, seen by all four surface antennas in the time domain. The signal of the event reconstructs to be coming from the building of the 10m telescope.}
 \label{fig08}
 \end{figure}

The reconstruction of the origin of RFI events seen in more than two antennas will indicate possible noise sources at the South Pole e.g. the IceCube counting house or the South Pole Station. A map of these sources will help to improve air shower detection.\\ 
A $\chi^2$ minimization on time residuals is used to reconstruct the source location of single events. To test the event reconstruction algorithm, we use signals generated with a GHz horn antenna in front of the MAPO building. Considering the cable delays and the antenna positions (Table~\ref{Table1}) we are able to make a 3D and time reconstruction of the events. Figure \ref{fig04} shows the reconstruction with horn antenna signal data is working well. It is accurate within several meters and shows clearly the horn antenna lies in front of the MAPO building near the antenna MAPO1. The horn antenna data reconstructs to the actual position within 50~m with an RMS of 2.3~m. Figure \ref{fig08} shows a typical noise event triggered with the RICE surface trigger. The event can be nicely reconstructed to have its origin in the building of the 10~m Telescope which is the topmost building in Fig.~\ref{fig03}.

\subsection{Events and background in the frequency domain}
Another Topic is the rate and variation of the different RFI sources during a whole year of measurement.
It is expected that RFI events have a typical signature in the frequency domain. This will help to find an ideal frequency region for a radio air shower detector. The continuous background is monitored over a whole year using the unbiased RICE events. One of the highest peaks on top of the continuous radio background is expected to be the meteor radar at 46.6~MHz to 47.0~MHz and 49.6~MHz to 50.0~MHz \cite{meteor}. It is clearly observable in the DFT of the recorded data together with a few other expected sources of filterable continuous narrow band RF signals.\\
The DFT of the constant background measured with the fat wire dipoles is the basis to evaluate the limit of detectable signal strength. For this it is of great importance to correct the measured data for the antenna properties, the high- and low pass filtering, the amplifier response, and the attenuation of the signal cable. Another important topic will be to determine long term variations of the background during one year.\\
RICE triggered surface events are studied in the frequency domain whether a discrimination of air shower radio signals from RFI noise is feasible. Most of the narrow band noise events could e.g. be filtered in a future air shower detector system.

\section{Conclusion}
As a part of RICE the four antenna surface detection system for radio signals, is able to study the conditions of the radio background in the frequency range from 25-150~MHz and higher at the South Pole. The threshold trigger strategy together with RICE allows for the estimation of the amount of RFI noise and its sources on the IceCube site. An analysis of the signals in the frequency domain shall be used to develop strategies to suppress the false trigger rate of a radio air shower detector. Measurements of the continuous background and its variations are the basis to estimate the energy threshold of a radio air shower detector in different frequency bands. The RFI measurements of the surface antennas will help to understand the signals measured with in ice radio detection systems.


\begin{thebibliography}{99}
 	 \bibitem{GeoSync}T. Huege, H. Falcke, Astron. \& Astrophys. 412, 1934, (2003).
   \bibitem{IceCube}  \url{http://www.icecube.wisc.edu}.
   \bibitem{ICRC07} J. Auffenberg \etal ICRC arXiv:0708.3331 (2007).
   \bibitem{NEC2} \url{http://www.nec2.org}.
   \bibitem{meteor} E. M. Lau \etal Radio Sci. 41, RS4007, (2006).
   \bibitem{Arena08} J. Auffenberg \etal Arena, doi:10.1016/j.nima.2009.03.179 (2008).
   \bibitem{RICE} I. Kravchenko \etal Phys. Rev. D 73, 082002 (2006).
 \end{thebibliography}
\end{document}